\documentclass[twocolumn,showpacs,preprintnumbers,amsmath,amssymb]{revtex4}

\usepackage{graphicx}
\usepackage{dcolumn}
\usepackage{bm}
\usepackage{amssymb}
\usepackage{amsfonts}
\usepackage{amsmath}
\usepackage{latexsym}
\usepackage{dsfont}
\usepackage{multirow}
\usepackage{color}
\usepackage[mathscr]{eucal}

\definecolor{nv}{rgb}{0.1,0.1,0.6}
\definecolor{pr}{rgb}{0.2,0.1,0.5}
\definecolor{mg}{rgb}{0.4,0.0,0.4}

\newcommand{\nn}{\nonumber}

\newcommand{\beq}{\begin{equation}}
\newcommand{\eeq}{\end{equation}}
\newcommand{\beqy}{\begin{eqnarray}}
\newcommand{\eeqy}{\end{eqnarray}}
\newcommand{\beqyn}{\begin{eqnarray*}}
\newcommand{\eeqyn}{\end{eqnarray*}}

\newcommand{\bs}{\begin{slide}}
\newcommand{\es}{\end{slide}}
\newcommand{\bc}{\begin{center}}
\newcommand{\ec}{\end{center}}
\newcommand{\bmin}{\begin{minipage}}
\newcommand{\emin}{\end{minipage}}

\newcommand{\bi}{\begin{itemize}}
\newcommand{\ei}{\end{itemize}}

\newcommand{\bea}{\begin{eqnarray}}
\newcommand{\eea}{\end{eqnarray}}
\newcommand{\be}{\begin{equation}}
\newcommand{\ee}{\end{equation}}

\newcommand{\ud}{\mathrm{d}}

\newcommand{\barpsi}{\overline{\psi}}

\newlength\savedwidth

\newcommand\whline{\noalign{\global\savedwidth\arrayrulewidth
\global\arrayrulewidth 1pt}%
\hline
\noalign{\global\arrayrulewidth\savedwidth}}
\newcommand{\uvec}[1]{\boldsymbol{#1}}

\newcommand{\LRD}{\overset{\leftrightarrow}{D}\!\!\!\!\!\phantom{D}}

\begin{document}


\title{Spin-orbit correlations in the nucleon}

\author{C\'edric Lorc\'e}
\email{lorce@ipno.in2p3.fr;C.Lorce@ulg.ac.be}
\affiliation{IPNO, Universit\'e Paris-Sud, CNRS/IN2P3, 91406 Orsay, France}
\affiliation{IFPA,  AGO Department, Universit\'e de Li\` ege, Sart-Tilman, 4000 Li\`ege, Belgium}

\date{\today}

\begin{abstract}
We investigate the correlations between the quark spin and orbital angular momentum inside the nucleon. Similarly to the Ji relation, we show that these correlations can be expressed in terms of specific moments of measurable parton distributions. This provides a whole new piece of information about the partonic structure of the nucleon.
\end{abstract}

\maketitle

\section{Introduction}

One of the key questions in hadronic physics is to unravel the spin structure of the nucleon, a very interesting playground for understanding many non-pertubative aspects of quantum chromodynamics (QCD). So far, most of the efforts have focused on the proper decomposition of the nucleon spin into quark/gluon and spin/orbital angular momentum (OAM) contributions (see Ref. \cite{Leader:2013jra} for a detailed recent review) and their experimental extraction. The spin structure is however richer than this.

Since the spin and OAM have negative intrinsic parity, the only non-vanishing single-parton ($a=q,G$) longitudinal correlations allowed by parity invariance are $\langle S^a_zS^N_z\rangle$, $\langle L^a_zS^N_z\rangle$ and $\langle L^a_zS^a_z\rangle$, where $\langle\quad\rangle$ denotes the appropriate average, $S^{q,G}_z$ is the quark/gluon longitudinal spin, $L^{q,G}_z$ is the quark/gluon longitudinal OAM and $S^N_z$ is the nucleon longitudinal spin. Since we are interested in the intrinsic correlations only, the global orbital motion of the system $L^N_z$ is not considered. The first two kinds of correlation are usually just called spin and OAM contributions of parton $a$ to the nucleon spin. The last type is simply the parton spin-orbit correlation. 

Even though generalized parton distributions (GPDs) and transverse-momentum dependent parton distributions (TMDs) are naturally sensitive to the parton spin-orbit correlations, no quantitative relation between them has been derived so far. The only quantitative relation we are aware of has been obtained in Ref.~\cite{Lorce:2011kd} at the level of generalized TMDs (GTMDs)~\cite{Meissner:2009ww,Lorce:2013pza}, also known as unintegrated GPDs (uGPDs), which are unfortunately not yet related to any experimental process.

In this Letter we provide the relation between the quark spin-orbit correlation and measurable parton distributions. Our approach is similar to the one used in Ref.~\cite{Ji:1996ek} in the case of quark OAM, but this time in the parity-odd sector and with an asymmetric tensor. The Letter is organized as follows: In section~\ref{sec2}, we define the quark spin-orbit correlation operator and express the corresponding expectation value in terms of form factors. In section~\ref{sec3} we relate these form factors to moments of measurable parton distributions. In section~\ref{sec4}, we provide an estimate of the various contributions, and conclude the paper with section~\ref{sec5}.

\section{Quark spin-orbit correlation}\label{sec2}

It is well known that the light-front operator giving the total number of quarks can be decomposed into the \emph{sum} of right- and left-handed quark contributions
\begin{align}
\hat N^q&=\int\ud^3x\,\barpsi \gamma^+\psi\\
&=\underbrace{\int\ud^3x\,\barpsi_R \gamma^+\psi_R}_{\hat N^{q_R}}+\underbrace{\int\ud^3x\,\barpsi_L \gamma^+\psi_L}_{\hat N^{q_L}},
\end{align}
where $\psi_{R,L}=\tfrac{1}{2}(\mathds 1\pm\gamma_5)\psi$, $a^\pm=\tfrac{1}{\sqrt{2}}(a^0+a^3)$ for a generic four-vector $a$, and $\ud^3x=\ud x^-\,\ud^2x_\perp$. The quark longitudinal spin operator simply corresponds to half of the \emph{difference} between right- and left-handed quark numbers
\begin{align}
\hat S^q_z&=\int\ud^3x\,\tfrac{1}{2}\,\barpsi \gamma^+\gamma_5\psi\\
&=\tfrac{1}{2}(\hat N^{q_R}-\hat N^{q_L}).
\end{align}

Similarly, we decompose the local gauge-invariant light-front operator for the quark longitudinal OAM~\cite{Ji:1996ek} into the \emph{sum} of right- and left-handed quark contributions
\begin{align}
\hat L^q_z&=\int\ud^3x\,\tfrac{1}{2}\,\barpsi \gamma^+(\uvec x\times i\overset{\leftrightarrow}{\uvec D}\!\!\!\!\!\phantom{D})_z\psi\\
&=\hat L^{q_R}_z+\hat L^{q_L}_z,
\end{align}
where $\overset{\leftrightarrow}{\uvec D}\!\!\!\!\!\phantom{D}=\overset{\rightarrow}{\uvec\partial}\!\!\!\!\!\phantom{\partial}-\overset{\leftarrow}{\uvec\partial}\!\!\!\!\!\phantom{\partial}-2ig\uvec A$ is the symmetric covariant derivative, and $\hat L^{q_{R,L}}_z=\int\ud^3x\,\tfrac{1}{2}\,\barpsi_{R,L} \gamma^+(\uvec x\times i\overset{\leftrightarrow}{\uvec D}\!\!\!\!\!\phantom{D})_z\psi_{R,L}$. The \emph{difference} between these right- and left-handed quark contributions will be referred to as the quark longitudinal spin-orbit correlation operator which reads
\begin{align}
\hat C^q_z&=\int\ud^3x\,\tfrac{1}{2}\,\barpsi \gamma^+\gamma_5(\uvec x\times i\overset{\leftrightarrow}{\uvec D}\!\!\!\!\!\phantom{D})_z\psi\label{SO}\\
&=\hat L^{q_R}_z-\hat L^{q_L}_z.
\end{align}

The quark spin and OAM operators attracted a lot of attention because they enter the Ji decomposition of the total angular momentum operator in QCD~\cite{Ji:1996ek}
\beq
\hat J_z=\hat S^q_z+\hat L^q_z+\hat J^G_z.
\eeq
Though, as we have seen,  a complete characterization of the nucleon longitudinal spin structure requires us to go beyond this and to consider the quark number and spin-orbit correlation operators as well. Contrary to the quark number, the quark spin-orbit correlation defined by Eq.~\eqref{SO} has, to the best of our knowledge, never been studied so far. The purpose of this Letter is to fill this gap and to show that such a quantity is actually related to measurable quantities.
\newline

We basically follow the same strategy as Ji in Ref.~\cite{Ji:1996ek}, except for the fact that we directly consider the more general asymmetric gauge-invariant energy-momentum tensor instead of the symmetric gauge-invariant (or Belinfante) one. We postpone the discussion of this particular point to Section~\ref{sec4}. The quark OAM operator can then conveniently be expressed as follows
\beq\label{JiOAMdef}
\hat L^q_z=\int\ud^3x\,(x^1\hat T^{+2}_q-x^2\hat T^{+1}_q),
\eeq
where $\hat T^{\mu\nu}$ is the quark energy-momentum tensor operator~\cite{Leader:2013jra}
\begin{align}
\hat T^{\mu\nu}_q&=\tfrac{1}{2}\,\barpsi\gamma^\mu\, i\LRD^\nu\psi\\
&=\hat T^{\mu\nu}_{q_R}+\hat T^{\mu\nu}_{q_L}
\end{align}
with $\hat T^{\mu\nu}_{q_{R,L}}=\tfrac{1}{2}\,\barpsi_{R,L}\gamma^\mu\, i\LRD^\nu\psi_{R,L}$. Similarly, we rewrite the quark spin-orbit operator as
\beq\label{OAM5}
\hat C^q_z=\int\ud^3x\,(x^1\hat T^{+2}_{q5}-x^2\hat T^{+1}_{q5}),
\eeq
where $\hat T^{\mu\nu}_{q5}$ can be considered as the parity-odd partner of the quark energy-momentum tensor operator
\begin{align}
\hat T^{\mu\nu}_{q5}&=\tfrac{1}{2}\,\barpsi\gamma^\mu\gamma_5\, i\LRD^\nu\psi\\
&=\hat T^{\mu\nu}_{q_R}-\hat T^{\mu\nu}_{q_L}.
\end{align}

Just like in the case of the generic asymmetric energy-momentum tensor \cite{Bakker:2004ib,Leader:2013jra}, we find that the non-forward matrix elements of $\hat T^{\mu\nu}_{q5}$ can be parametrized in terms of five form factors (FFs)
\beq\label{EMTparam}
\langle p',\uvec s'|\hat T^{\mu\nu}_{q5}|p,\uvec s\rangle=\overline u(p',\uvec s')\Gamma^{\mu\nu}_{q5}u(p,\uvec s)
\eeq
with
\begin{align}
\Gamma^{\mu\nu}_{q5}&=\tfrac{P^{\{\mu}\gamma^{\nu\}}\gamma_5}{2}\,\tilde A_q(t)+\tfrac{P^{\{\mu}\Delta^{\nu\}}\gamma_5}{4M}\,\tilde B_q(t)\nn\\
&\quad+\tfrac{P^{[\mu}\gamma^{\nu]}\gamma_5}{2}\,\tilde C_q(t)+\tfrac{P^{[\mu}\Delta^{\nu]}\gamma_5}{4M}\,\tilde D_q(t)\nn\\
&\quad+Mi\sigma^{\mu\nu}\gamma_5\,\tilde F_q(t),
\end{align}
where $M$ is the nucleon mass, $\uvec s$ and $\uvec s'$ are the initial and final rest-frame polarization vectors satisfying $\uvec s^2=\uvec s'^2=1$, $P=\tfrac{p'+p}{2}$ is the average four-momentum, and $t=\Delta^2$ is the square of the four-momentum transfer $\Delta=p'-p$. For convenience, we used the notations $a^{\{\mu}b^{\nu\}}=a^\mu b^\nu+a^\nu b^\mu$ and $a^{[\mu}b^{\nu]}=a^\mu b^\nu-a^\nu b^\mu$. 

Since we are interested in the matrix element of Eq.~\eqref{OAM5}  which involves only one explicit power of $\uvec  x$, we need to expand Eq.~\eqref{EMTparam} only up to linear order in $\Delta$~\cite{Leader:2013jra,Bakker:2004ib}. Considering initial and final nucleon states with the same rest-frame polarization $\uvec s'=\uvec s=(\uvec s_\perp,s_z)$ and using the light-front spinors (see \emph{e.g.} Appendix A of Ref.~\cite{Lorce:2011zta}), we arrive at the following expression 
\begin{align}\label{expansion}
\langle p',\uvec s|\hat T^{\mu\nu}_{q5}|p,\uvec s\rangle&=\left[P^{\{\mu}S^{\nu\}}-\tfrac{P^{\{\mu}i\epsilon^{\nu\}+\Delta P}}{2P^+}\right]\tilde A_q\nn\\
&\,\quad\left[P^{[\mu}S^{\nu]}-\tfrac{P^{[\mu}i\epsilon^{\nu]+\Delta P}}{2P^+}\right](\tilde C_q-2\tilde F_q)\nn\\
&\,\quad+i\epsilon^{\mu\nu\Delta P}\,\tilde F_q+\mathcal O(\Delta^2)
\end{align}
with $\epsilon_{0123}=+1$ and the covariant spin vector $S^\mu=[s_zP^+,-s_zP^-+\tfrac{\uvec P_\perp}{P^+}\cdot(M\uvec s_\perp+\uvec P_\perp s_z),M\uvec s_\perp+\uvec P_\perp s_z]$ satisfying $P\cdot S=0$ and $S^2=-M^2-s^2_z(P^2-M^2)$. For convenience, we removed the argument of the FFs when evaluated at $t=0$, \emph{i.e.} $\tilde X_q=\tilde X_q(0)$. 

Substituting the expansion~\eqref{expansion} into the matrix element of Eq.~\eqref{OAM5} and working in the symmetric light-front frame, \emph{i.e.} with $\uvec P_\perp=\uvec 0_\perp$, we find
\beq
C^q_z\equiv\tfrac{\langle P,\uvec e_z|\hat C^q_z|P,\uvec e_z\rangle}{\langle P,\uvec e_z|P,\uvec e_z\rangle}=\tfrac{1}{2}(\tilde A_q+\tilde C_q).
\eeq
Thus, to determine the quark spin-orbit correlation, one has to measure the $\tilde A_q(t)$ and $\tilde C_q(t)$ FFs, which are analogous to the axial-vector FF $G^q_A(t)$. The $\tilde B_q(t)$ and $\tilde D_q(t)$ FFs, which are analogous to the induced pseudoscalar FF $G^q_P(t)$, are not needed since they contribute only to higher $x$-moments of $\hat T^{\mu\nu}_{q5}$, as one can see from the expansion $\overline u(p',\uvec s)\,\tfrac{P^\mu\Delta^\nu\gamma_5}{4M}\,u(p,\uvec s)=\mathcal O(\Delta^2)$.

\section{Link with parton distributions}\label{sec3}

Like in the case of the energy-momentum tensor, there is no fundamental probe that couples to $\hat T^{\mu\nu}_{q5}$ in particle physics. However, it is possible to relate the corresponding various FFs to specific moments of measurable parton distributions. From the component $\hat T^{++}_{q5}$, we find
\begin{align}
\int\ud x\, x\tilde H_q(x,\xi,t)&=\tilde A_q(t),\\
\int\ud x\, x\tilde E_q(x,\xi,t)&=\tilde B_q(t),
\end{align}
where $\tilde H_q(x,\xi,t)$ and $\tilde E_q(x,\xi,t)$ are the GPDs parametrizing the non-local twist-2 axial-vector light-front quark correlator~\cite{Diehl:2003ny}
\begin{align}
\frac{1}{2}\int\frac{\ud z^-}{2\pi}\,e^{ixP^+z-}&\langle p',\uvec s'|\barpsi(-\tfrac{z^-}{2})\gamma^+\gamma_5\mathcal W\psi(\tfrac{z^-}{2})|p,\uvec s\rangle\nn\\
&=\tfrac{1}{2P^+}\,\overline u(p',\uvec s')\Gamma^+_{qA}u(p,\uvec s)
\end{align}
with $\mathcal W=\mathcal P \exp[ ig\int^{-z^-/2}_{z^-/2}\ud y^-A^+(y^-)]$ a straight light-front Wilson line and
\beq
\Gamma^+_{qA}=\gamma^+\gamma_5\,\tilde H_q(x,\xi,t)+\tfrac{\Delta^+\gamma_5}{2M}\,\tilde E_q(x,\xi,t),
\eeq
the skewness variable being given by $\xi=-\Delta^+/2P^+$. 

The relations for the other FFs can be obtained using the following QCD identity
\beq\label{QCDidentity}
\barpsi\gamma^{[\mu}\gamma_5\, i\LRD^{\nu]}\psi=2m\,\barpsi i\sigma^{\mu\nu}\gamma_5\psi-\epsilon^{\mu\nu\alpha\beta}\partial_\alpha(\barpsi\gamma_\beta\psi),
\eeq
where $m$ is the quark mass. Taking the matrix elements of both sides and using some Gordon and  $\epsilon$-identities, we find
\begin{align}
\tilde C_q(t)&=\tfrac{m}{2M}\,H^q_1(t)-F^q_1(t),\label{Cq}\\
\tilde D_q(t)&=\tfrac{m}{2M}\,H^q_2(t)-F^q_2(t),\\
\tilde F_q(t)&=\tfrac{m}{2M}\,H^q_3(t)-\tfrac{1}{2}\,G^q_E(t),\label{Fq}
\end{align}
where the electric Sachs FF is given by $G^q_E(t)=F^q_1(t)+\tfrac{t}{4M^2}\,F^q_2(t)$. The FFs on the right-hand side parametrize the vector and tensor local correlators as follows
\begin{align}
\langle p',\uvec s'|\barpsi\gamma^\mu\psi|p,\uvec s\rangle&=\overline u(p',\uvec s')\Gamma^\mu_{qV}u(p,\uvec s),\\
\langle p',\uvec s'|\barpsi i\sigma^{\mu\nu}\gamma_5\psi|p,\uvec s\rangle&=\overline u(p',\uvec s')\Gamma^{\mu\nu}_{qT}u(p,\uvec s)
\end{align}
with
\begin{align}
\Gamma^\mu_{qV}&=\gamma^\mu F^q_1(t)+\tfrac{i\sigma^{\mu\nu}\Delta_\nu}{2M}\,F^q_2(t),\\
\Gamma^{\mu\nu}_{qT}&=\tfrac{P^{[\mu}\gamma^{\nu]}\gamma_5}{2M}\,H^q_1(t)+\tfrac{P^{[\mu}\Delta^{\nu]}\gamma_5}{4M^2}\,H^q_2(t)\nn\\
&\quad+i\sigma^{\mu\nu}\gamma_5\,H^q_3(t).
\end{align}
Our tensor FFs are related to the ones used in Refs.~\cite{Hagler:2004yt,Diehl:2005jf} in the following way
\begin{align}
A^q_{T10}(t)&=-\tfrac{1}{2}[H^q_1(t)+\tfrac{t}{4M^2}\,H^q_2(t)]+H^q_3(t),\\
B^q_{T10}(t)&=\tfrac{1}{2}[H^q_1(t)+H^q_2(t)],\\
\tilde A^q_{T10}(t)&=-\tfrac{1}{4}\,H^q_2(t),
\end{align}
and can also be expressed in terms of moments of twist-2 chiral-odd GPDs.

The quark spin-orbit correlation is therefore given by the simple expression
\beq\label{SOtwist2}
C^q_z=\tfrac{1}{2}\int\ud x\,x\tilde H_q(x,0,0)-\tfrac{1}{2}\,[F^q_1(0)-\tfrac{m}{2M}\,H^q_1(0)]
\eeq
which is the analogue in the parity-odd sector of the Ji relation~\cite{Ji:1996ek} for the quark OAM
\beq\label{Jirel}
L^q_z=\tfrac{1}{2}\int\ud x\,x[H_q(x,0,0)+E_q(x,0,0)]-\tfrac{1}{2}\,G^q_A(0).
\eeq
It may seem counter-intuitive that the spin-orbit correlation $C^q_z$ is related to the spin-spin correlation $\tilde H_q$. But, just like in the Ji relation, the ``orbital'' information is actually provided by the extra $x$-factor representing the fraction of longitudinal momentum, as discussed by Burkardt in Ref.~\cite{Burkardt:2005hp}.

From the components $\hat T^{j+}_{q5}$ with $j=1,2$ we also find the following relations
\begin{align}
\int\ud x\,x\tilde G^q_1(x,\xi,t)&=-\tfrac{1}{2}\,[\tilde B_q(t)+\tilde D_q(t)],\label{tildeG1}\\
\int\ud x\,x\tilde G^q_2(x,\xi,t)&=-\tfrac{1}{2}\,[\tilde A_q(t)+\tilde C_q(t)]+(1-\xi^2)\tilde F_q(t),\\
\int\ud x\,x\tilde G^q_3(x,\xi,t)&=-\tfrac{\xi}{2}\,\tilde F_q(t),\\
\int\ud x\,x\tilde G^q_4(x,\xi,t)&=-\tfrac{1}{2}\,\tilde F_q(t),\label{tildeG2}
\end{align}
where $\tilde G^q_i(x,\xi,t)$ are the GPDs parametrizing the non-local twist-3 axial-vector light-front quark correlator~\cite{Kiptily:2002nx}. As a check, we inserted in Eqs.~\eqref{tildeG1}-\eqref{tildeG2} the expressions for $\int\ud x\,x\tilde G^q_i(x,\xi,t)$ derived in Ref.~\cite{Kiptily:2002nx} within the Wandzura-Wilczek approximation (which is exact for the lowest two $x$-moments in the chiral limit~\cite{Kiptily:2002nx}) and consistently recovered Eqs.~\eqref{Cq}-\eqref{Fq} in the massless quark limit. The analogue of the Penttinen-Polyakov-Shuvaev-Strikman relation~\cite{Kiptily:2002nx,Penttinen:2000dg}
\beq
L^q_z=-\int\ud x\,xG_2(x,0,0)
\eeq
 in the parity-odd sector is therefore
\beq\label{SOtwist3}
C^q_z=-\int\ud x\,x[\tilde G^q_2(x,0,0)+2\tilde G^q_4(x,0,0)].
\eeq

For completeness, we note that the quark spin-orbit correlation can also be expressed in terms of GTMDs
\beq\label{SOGTMDs}
C^q_z=\int\ud x\,\ud^2k_\perp\,\tfrac{\uvec k^2_\perp}{M^2}\,G^q_{11}(x,0,\uvec k^2_\perp,0,0).
\eeq
As discussed in Refs.~\cite{Hatta:2011ku,Ji:2012sj,Lorce:2012ce,Burkardt:2012sd}, the shape of the Wilson line entering the definition of the GTMDs determines the type of OAM. Since in this Letter we are interested in the Ji OAM, the GTMD $G^q_{11}(x,\xi,\uvec k^2_\perp,\uvec k_\perp\cdot\uvec \Delta_\perp,\uvec\Delta_\perp^2)$ in Eq.~\eqref{SOGTMDs} has to be defined with a direct straight Wilson line. On the contrary, the spin-orbit correlation introduced in Ref.~\cite{Lorce:2011kd} dealt with the Jaffe-Manohar OAM which is obtained by defining the GTMDs with a staple-like light-front Wilson line.

\section{Discussion}\label{sec4}

In the previous section, we have obtained three different expressions for the quark spin-orbit correlation in terms of parton distributions. From an experimental point of view, Eq.~\eqref{SOtwist2} is clearly the most useful one. By equating the right-hand side of Eq.~\eqref{SOtwist2} with the right-hand sides of Eqs.~\eqref{SOtwist3} and \eqref{SOGTMDs}, we obtain two new sum rules among parton distributions.

In order to determine the quark spin-orbit correlation, we need to know three quantities. The first quantity is the Dirac FF evaluated at $t=0$ which simply corresponds to the valence number, namely $F^u_1(0)=2$ and $F^d_1(0)=1$ in a proton, and therefore does not require any experimental input. The second quantity is the tensor FF $H^q_1(0)$ which is not known so far. However, since it appears multiplied by the mass ratio $m/4M\sim 10^{-3}$ for $u$ and $d$ quarks in Eq.~\eqref{SOtwist2}, we do not expect it to contribute significantly to $C^q_z$. The last quantity is the second moment of the quark helicity distribution
\beq
\int_{-1}^1\ud x\,x\tilde H_q(x,0,0)=\int_0^1\ud x\,x[\Delta q(x)-\Delta\overline q(x)].
\eeq
Contrary to the lowest moment $\int_{-1}^1\ud x\,\tilde H_q(x,0,0)=\int_0^1\ud x\,[\Delta q(x)+\Delta\overline q(x)]$, this second moment cannot be extracted from deep-inelastic scattering (DIS) polarized data without additional assumptions about the polarized sea quark densities. The separate quark and antiquark contributions can however be obtained \emph{e.g.} in a combined fit to inclusive and semi-inclusive DIS. From the Leader-Sidorov-Stamenov (LSS) analysis of Ref.~\cite{Leader:2010rb}, we obtain 
\begin{align}
\int_{-1}^1\ud x\,x\tilde H_u(x,0,0)&\approx 0.19,\\
\int_{-1}^1\ud x\,x\tilde H_d(x,0,0)&\approx -0.06,
\end{align}
at the scale $\mu^2=1$ GeV$^2$, leading to $C^u_z\approx -0.9$ and $C^d_z\approx -0.53$ . Note that these estimates are however not particularly reliable since it follows from the new HERMES~\cite{Airapetian:2012ki} and COMPASS~\cite{Makke:2013bya} data on multiplicities that the fragmentation functions given in Ref.~\cite{de Florian:2007hc} and used in the LSS analysis are presumably not correct~\cite{Leader:2013kra}. Further experimental data and dedicated analyses are therefore required. Nevertheless, these values seem consistent with recent Lattice calculations by the LHPC collaboration~\cite{Bratt:2010jn}, see table~\ref{Modelresults}.

The second moment of the quark helicity distribution being a valence-like quantity with suppressed low-$x$ region, we may expect phenomenological quark model predictions to be more reliable for this second moment than for the lowest one. In table~\ref{Modelresults} we provide the first two moments of the $u$ and $d$-quark helicity distributions obtained within the naive quark model (NQM), the light-front constituent quark model (LFCQM) and the light-front chiral quark-soliton model (LF$\chi$QSM) at the scale $\mu^2\sim 0.26$ GeV$^2$, see Ref.~\cite{Lorce:2011dv} for more details. From these estimates, we expect a negative quark spin-orbit $C^q_z$ for both $u$ and $d$ quarks ($C^u_z\approx -0.8$ and $C^d_z\approx -0.55$), meaning that the quark spin and Ji OAM are expected to be, in average, anti-correlated. This has to be contrasted with the model results obtained in Ref.~\cite{Lorce:2011kd} where the quark spin and Jaffe-Manohar OAM are, in average, correlated.
\newline

\begin{table}[t!]
\begin{center}
\caption{\footnotesize{Comparison between the lowest two axial moments $\Delta^{(n)}q\equiv\int^1_{-1}\ud x\,x^n\tilde H_q(x,0,0)$ for $q=u,d$ as predicted by the naive quark model (NQM), the light-front constituent quark model (LFCQM) and the light-front chiral quark-soliton model (LF$\chi$QSM) at the scale $\mu^2\sim 0.26$ GeV$^2$~\cite{Lorce:2011dv}, with the corresponding values obtained from the LSS fit to experimental data~\cite{Leader:2010rb} at the scale $\mu^2=1$ GeV$^2$ and Lattice calculations~\cite{Bratt:2010jn} at the scale $\mu^2=4$ GeV$^2$ and pion mass $m_\pi=293$ MeV.}}\label{Modelresults}
\begin{tabular}{c@{\quad}|@{\quad}c@{\quad}c@{\quad}c@{\quad}c}\whline
Model&$\Delta^{(0)} u$&$\Delta^{(0)} d$&$\Delta^{(1)}u$&$\Delta^{(1)}d$\\
\hline
NQM&$4/3$&$-1/3$&$4/9$&$-1/9$\\
LFCQM&$0.995$&$-0.249$&$0.345$&$-0.086$\\
LF$\chi$QSM&$1.148$&$-0.287$&$0.392$&$-0.098$\\
\hline
Exp.&$0.82$&$-0.45$&$\approx 0.19$&$\approx -0.06$\\
Latt.&$0.82(7)$&$-0.41(7)$&$\approx 0.20$&$\approx -0.05$\\
\whline
\end{tabular}
\end{center}
\end{table}

Finally, we would like to comment about the difference between symmetric and asymmetric tensors. The canonical energy-momentum tensor obtained from Noether theorem is conserved, but is usually neither gauge-invariant nor symmetric. One has however the freedom to define alternative energy-momentum tensors by adding so-called superpotential terms~\cite{Leader:2013jra,Ji:1996ek,Jaffe:1989jz} to the canonical energy-momentum tensor. These terms are of the form $\partial_\lambda G^{\lambda\mu\nu}$ with $G^{\mu\lambda\nu}=-G^{\lambda\mu\nu}$, so that the new tensors remain conserved and the \emph{total} four-momentum (\emph{i.e.} integrated over all space) remains unchanged. Physically, adding superpotential terms corresponds to relocalizing/redefining what we mean by density of energy and momentum~\cite{Hehl:1976vr}. 

The Belinfante procedure simply makes use of this freedom to define a symmetric and gauge-invariant tensor. It is however important to realize that the symmetry property of the energy-momentum tensor is not mandatory in particle physics, but is basically motivated by General Relativity where gravitation is coupled to a symmetric energy-momentum tensor. From the conservation of total angular momentum $\partial_\mu\left(x^\nu T^{\mu\rho}-x^\rho T^{\mu\nu}+S^{\mu\nu\rho}\right)=0$, one sees that the antisymmetric part of the energy-momentum tensor is intimately related to the spin density
\beq
T^{\nu\rho}-T^{\rho\nu}=-\partial_\mu S^{\mu\nu\rho}. 
\eeq
General Relativity is a purely classical theory where the notion of intrinsic spin does not exist, and so it is hardly surprising that only a symmetric energy-momentum tensor is needed in that context. 

So, as soon as one deals with a spin density, the natural gauge-invariant energy-momentum tensor will be asymmetric. The fact that one can still define a consistent symmetric (or Belinfante) tensor is a consequence of the following QCD identity
\beq\label{QCDidentity2}
\barpsi\gamma^{[\mu} i\LRD^{\nu]}\psi=-\epsilon^{\mu\nu\alpha\beta}\partial_\alpha(\barpsi\gamma_\beta\gamma_5\psi),
\eeq
which is the parity-even analogue of Eq.~\eqref{QCDidentity}. It tells us that the difference between the symmetric and asymmetric tensors (namely the antisymmetric part) is nothing but a superpotential term. This means that one can effectively absorb the contribution of the spin density in a relocalization/redefinition of energy-momentum density.

In Ref.~\cite{Ji:1996ek}, Ji considered the symmetric (Belinfante) energy-momentum tensor, discarding from the very beginning the notion of intrinsic spin density. Accordingly, Ji was able to relate only the \emph{total} angular momentum to moments of GPDs
\beq
J^{q,G}_z=\tfrac{1}{2}\int\ud x\,x[H_{q,G}(x,0,0)+E_{q,G}(x,0,0)].
\eeq
In order to get the quark OAM, one has to subtract as a second step the quark spin contribution $S^q_z=\tfrac{1}{2}\,G^q_A(0)$ from the total quark angular momentum contribution $J^q_z$. So Eq.~\eqref{Jirel} can be understood as $L^q_z=J^q_z-S^q_z$, or more precisely as $\langle L^q_zS^N_z\rangle=\langle J^q_zS^N_z\rangle-\langle S^q_zS^N_z\rangle$. Starting with the more general asymmetric energy-momentum tensor, one can directly define the OAM operator as in Eq.~\eqref{JiOAMdef} and get the final result~\eqref{Jirel} without invoking any additional step~\cite{Leader:2013jra,Bakker:2004ib,Shore:1999be}.

Similarly, one can define another quark correlation using only the symmetric part of $\hat T^{\mu\nu}_{q5}$ in close analogy with Eq.~\eqref{OAM5}
\begin{align}
\hat{\mathds C}^q_z&=\int\ud^3x\,(x^1\tfrac{1}{2}\,\hat T^{\{+2\}}_{q5}-x^2\tfrac{1}{2}\,\hat T^{\{+1\}}_{q5})\\
&=\tfrac{1}{2}\,\hat C^q_z+\int\ud^3x\,\tfrac{1}{4}\,\barpsi i\overset{\leftrightarrow}{D}\!\!\!\!\!\phantom{D}^+(\uvec x\times\uvec\gamma )_z\gamma_5\psi.\nonumber
\end{align}
Contrary to $\hat C^q_z$, the operator $\hat{\mathds C}^q_z$ is time-independent since the corresponding current $\hat M^{\mu\nu\rho}_{q5}=x^\nu \tfrac{1}{2}\,\hat T^{\{\mu\rho\}}_{q5}-x^\rho\tfrac{1}{2}\,\hat T^{\{\mu\nu\}}_{q5}$ is conserved $\partial_\mu\hat M^{\mu\nu\rho}_{q5}=0$, as a consequence of the symmetry of $\hat T^{\{\mu\nu\}}_{q5}$. Note however that even if the operator $\hat C^q_z$ is in general time-dependent, its expectation value is time-\emph{in}dependent because the initial and final nucleon states have the same energy~\cite{Leader:2013jra}. 

It is pretty simple now to get the expressions for the matrix element of $\hat{\mathds C}^q_z$. We need to keep only the symmetric part of Eq.~\eqref{EMTparam}, \emph{i.e.} set the FFs $\tilde C_q$, $\tilde D_q$ and $\tilde F_q$ to zero in the derivation of Section \ref{sec2}. We then find 
\begin{align}
\mathds C^q_z\equiv\tfrac{\langle P,\uvec e_z|\hat{\mathds C}^q_z|P,\uvec e_z\rangle}{\langle P,\uvec e_z|P,\uvec e_z\rangle}&=\tfrac{1}{2}\,\tilde A_q(0),\\
&=\tfrac{1}{2}\int_{-1}^1\ud x\,x\tilde H_q(x,0,0).
\end{align}
Contrary to $C^q_z$, the physical interpretation of $\mathds C^q_z$ is not particularly clear owing to the mass term in Eq.~\eqref{QCDidentity}. Indeed, this mass term does not have the form of a superpotential, which in turn implies that $\int\ud^3x\,\hat T^{+\nu}_{q5}\neq\int\ud^3x\,\tfrac{1}{2}\,\hat T^{\{+\nu\}}_{q5}$. We argued in section~\ref{sec2} that $C^q_z$ can naturally be interpreted as (twice) the quark spin-orbit correlation $\langle L^q_zS^q_z\rangle$. By analogy with the parity-even sector, it is tempting to interpret $\mathds C^q_z$ as (twice) the correlation between quark spin and quark total angular momentum $\langle J^q_zS^q_z\rangle$. This is actually true if one gets rid of the mass term, \emph{i.e.} work in the chiral limit $m=0$. In this case, just like in the parity-even sector with the Ji relation, we can understand Eq.~\eqref{SOtwist2} as $\langle L^q_zS^q_z\rangle=\langle J^q_zS^q_z\rangle-\langle S^q_zS^q_z\rangle$. The quark spin-spin correlation $\langle S^q_zS^q_z\rangle$ is of course quite trivial and does not depend on the spin orientation. That is the reason why the unpolarized quark FF $F^q_1(0)$ appears in the expression for $C^q_z$.


\section{Conclusions}\label{sec5}

We provided a local gauge-invariant definition of the quark spin-orbit correlation, which is a new independent piece of information about the nucleon longitudinal spin structure. We derived several expressions for the expectation value of this correlation in terms of measurable parton distributions, leading to new sum rules. Using estimates from fits to available experimental data, Lattice calculations and phenomenological quark models, we concluded that the quark spin-orbit correlation is very likely negative, meaning that the quark spin and Ji orbital angular momentum are, in average, anti-correlated. However, a more precise determination of this quark spin-orbit correlation requires further experimental data and dedicated analyses in order to disentangle quark and antiquark contributions to the helicity distribution. Tensor form factors are also in principle needed, but the corresponding contribution can safely be neglected for light quarks.

\section*{Acknowledgements}

I am thankful to H.~Moutarde, B.~Pasquini and M.~Polyakov for useful discussions related to this study. I am also very grateful to E.~Leader, A.~V.~Sidorov and D.~B.~Stamenov for providing me with their extraction of the quark helicity distributions, and to M.~Engelhardt for informing me about recent Lattice results. This work was supported by the Belgian Fund F.R.S.-FNRS \emph{via} the contract of Charg\'e de recherches.

\end{document}